# On-chip two-octave supercontinuum generation by enhancing self-steepening of optical pulses

Lin Zhang[1], Yan Yan[1], Yang Yue[1], Qiang Lin[2], Oskar Painter[2], Raymond G. Beausoleil[3], and Alan E. Willner[1]

[1]Department of Electrical Engineering, University of Southern California, Los Angeles, CA 90089, USA
[2] Thomas J. Watson, Sr., Laboratory of Applied Physics, California Institute of Technology, Pasadena, CA 91125
[3] HP Laboratories, Palo Alto, CA 94304, USA

linzhang@usc.edu

**Abstract:** Dramatic advances in supercontinuum generation have been made recently using photonic crystal fibers, but it is quite challenging to obtain an octave-spanning supercontinuum on a chip, partially because of strong dispersion in high-index-contrast nonlinear integrated waveguides. We show by simulation that extremely flat and low dispersion can be achieved in silicon nitride slot waveguides over a wavelength band of 500 nm. Different from previously reported supercontinua that were generated either by higher-order soliton fission in anomalous dispersion regime or by self phase modulation in normal dispersion regime, a two-octave supercontinuum from 630 to 2650 nm (360 THz in total) can be generated by greatly enhancing self-steepening in nonlinear pulse propagation in almost zero dispersion regime, when an optical shock as short as 3 fs is formed, which enables on-chip ultra-wide-band applications.

**Introduction**

One basic building block in nonlinear optics is a supercontinuum generator, which has experienced a revolutionary development after its realization using photonic crystal fibers (PCFs) [1, 2]. Supercontinua of a few octaves in width have been reported [3, 4] for numerous applications such as frequency metrology, optical coherence tomography, pulse compression, microscopy and spectroscopy, telecommunication, and sensing. A key figure of merit is the width of a supercontinuum, which is greatly affected by the spectral profile of the dispersion in a nonlinear medium. The success of PCF-based supercontinuum generation is partially attributed to advanced dispersion engineering allowed by design freedom of the 2D lattice in the fiber cladding [2, 5]. Generally, the dispersion engineering is aimed at desirable zero-dispersion wavelengths and low dispersion over a wide spectral band. Flat dispersion of ±2 ps/(nm·km) over a 1000-nm-wide wavelength range was proposed [6, 7].

Highly nonlinear integrated waveguides and photonic wires with high index contrast have generated much excitement in recent years [8, 9], forming the backbone of compact devices in a photonic-integrated-circuit platform. However, to the best of our knowledge, demonstrated on-chip supercontinua have a spectral range of ~400 nm [10-14], far less than one octave, which is partially because of insufficient capability to engineer the dispersion of nonlinear waveguides. Recently, the dispersion profile of a silicon waveguide was made 20 times flatter by introducing a nano-scale slot structure [15], but this is still not sufficient to support more than one octave spectral broadening of femtosecond optical pulses.

Here, we propose a silicon nitride slot waveguide, which exhibits further 30× improvement in dispersion flatness, compared with that reported in Ref. [15]. In our simulation, a two-octave supercontinuum can be obtained on a chip by enhancing pulse self-steepening and forming an optical shock as short as 3 fs. The supercontinuum can be 'transferred' to third-harmonic spectral range by phase-locked pulse trapping effect. On-chip supercontinuum generation is believed to be a key enabler for building portable imaging, sensing, and frequency-metrology-based positioning systems. The advanced dispersion engineering technique in integrated waveguides is expected to open the door to combine ultrafast optics and nano-photonics and to apply ultra-wideband optical information technologies ubiquitously.

**Flat all-normal dispersion in silicon nitride slot waveguide**

The proposed silicon nitride ($Si_3N_4$) slot waveguide is shown in Fig. 1. A horizontal silica slot is formed between two $Si_3N_4$ layers. The substrate is 2-μm-thick $SiO_2$. The waveguide parameters are: width $W$ = 980 nm, upper height $H_u$ = 497 nm, lower height $H_l$ = 880 nm, and slot height $H_s$ = 120.5 nm. Material dispersions of $Si_3N_4$ and $SiO_2$ are considered. The waveguide has a single fundamental mode at the vertical polarization beyond 1800 nm.

Figure 2 shows the flat dispersion of quasi-TM (vertically polarized) mode in the slot waveguide in terms of dispersion coefficient $\beta_2 = d^2\beta(\omega)/d\omega^2$, where $\beta(\omega)$ is propagation constant. The average $\beta_2$ and $\Delta\beta_2$ are 0.0137 and 0.00195 $ps^2/m$, respectively, over a 610-nm-wide band from 1210 to 1820 nm. Compared to a dispersion-flattened silicon slot waveguide [15], which has $\beta_2$ of ±0.024 $ps^2/m$ from 1565 to 2100 nm, the proposed slot waveguide shows 30 times flatter dispersion. We intentionally modify the slot waveguide to obtain normal dispersion at all wavelengths because the supercontinuum generated in all-normal dispersion regime typically would have good spectral coherence [2, 16]. The dispersion flattening results from an anti-crossing effect [15]. The strip mode at short wavelength becomes more like a slot mode at long wavelength, which causes a slightly negative waveguide dispersion [15] to balance the convex dispersion in strip waveguides without a slot structure. The negative dispersion is made well-matched to the convex dispersion when $Si_3N_4$ and $SiO_2$ are adopted. Thus we obtain a much flatter dispersion than where using $Si/SiO_2$ slot waveguides. One can tailor dispersion value and slope for various nonlinear applications by changing waveguide structural parameters with similar trends presented in Ref. [15].

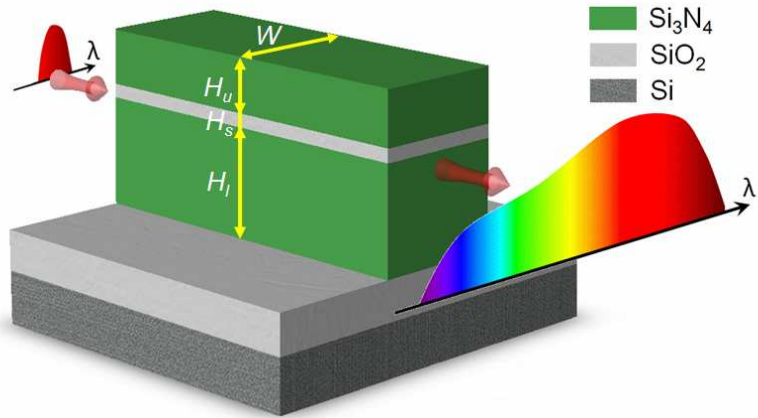

Fig. 1 Silicon nitride slot waveguide for dispersion flattening and supercontinuum generation. A horizontal silica slot is between two silicon nitride layers.

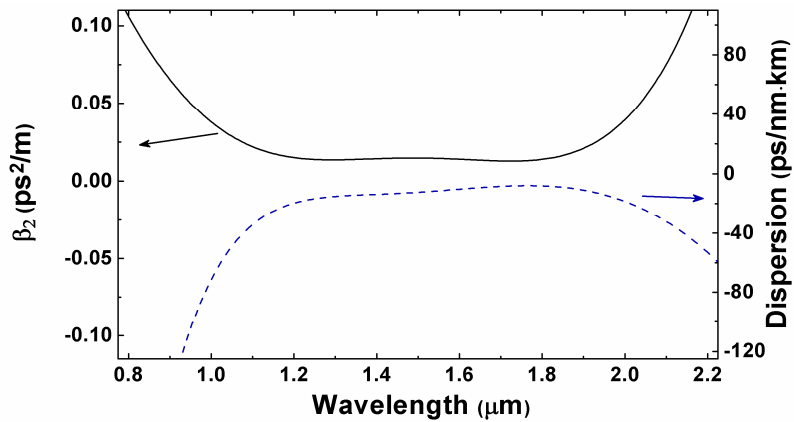

Fig. 2 Flattened all normal dispersion for supercontinuum generation in the slot waveguide.

**Supercontinuum generation by enhanced nonlinear self-steepening effect**

We use a generalized nonlinear envelope equation (GNEE) [17, 18], with third harmonic generation considered, to model supercontinuum generation. It has been confirmed that the simulation of even sub-cycle pulse propagations using this envelope equation is quite accurate [17-20], which is in excellent quantitative agreement with numerical integration of accurate Maxwell's equations [21]. We use split-step Fourier method [22] to solve the GNEE.

The propagation loss is set to be 1 dB/cm, which is achievable [23], and nonlinear loss induced by two-photon absorption is ignored [23]. All order dispersion terms are included. The nonlinear index $n_2$ for $Si_3N_4$ is measured at 1550 nm [24]. The Kerr nonlinear coefficient $\gamma_e$ is computed versus wavelength using the full-vectorial model [25]. The Raman scattering properties are considered [26]. We take the wavelength-dependence of the nonlinearity into account by correcting the shock terms.

A chirp-free hyperbolic secant pulse spectrally centered at 2200-nm wavelength, with a full width at half-maximum (FWHM) of 120 fs and a peak power of 6 kW (pulse energy of 0.8 nJ), is launched into the proposed $Si_3N_4$ slot waveguide. Figure 3(a) shows that, along the waveguide, the pulse generates significantly blue-shifted spectral components down to 800 nm wavelength mainly due to self-steepening. In our case, the flattened dispersion reduces walk-off of newly generated spectral components, which facilitates the formation of an optical shock at the pulse falling edge that is as short as 3 fs (see Fig. 3(b)), as it travels 5 mm. The optical shock induces so much spectral blue shift that it reaches a short-wavelength region where third harmonics are generated. The blue part of the spectrum becomes stable at a propagation distance of 15 mm and shows a small power fluctuation of 3 dB over a 754-nm-wide wavelength range from 847 to 1601 nm at 20 mm. The high-power part of the spectrum is red-shifted and extended to 2650 nm, due to self-phase modulation and Raman self-frequency shift [22]. The supercontinuum is formed from 630 to 2650 nm (that is, 360 THz in total) at –35 dB, covering a two-octave bandwidth.

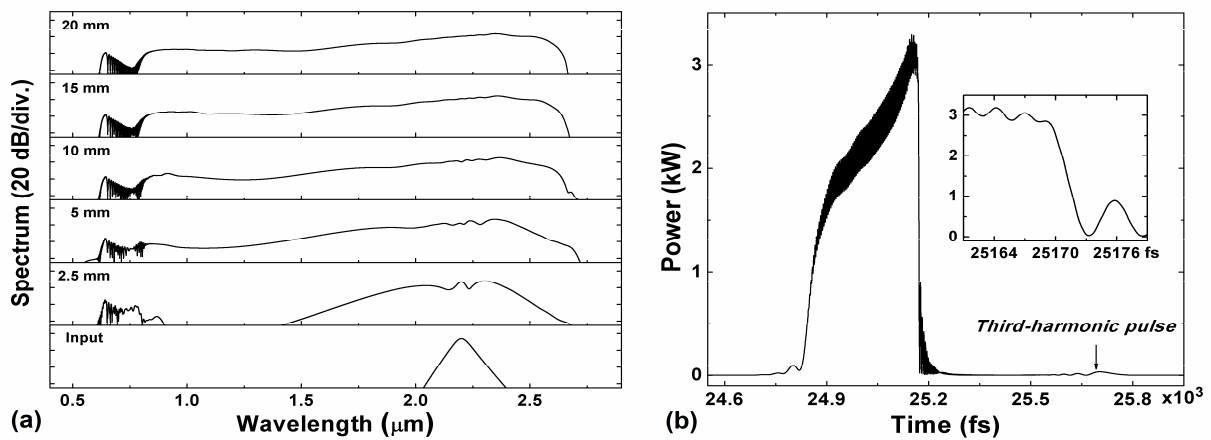

Fig. 3. Two-octave supercontinuum generation in the dispersion-flattened slot waveguide. (a) The input 120-fs pulse is centered at 2200-nm wavelength. A supercontinuum is generated from 630 to 2650 nm mainly due to self-steepening of the pulse. (b) In time domain, an optical shock as short as 3 fs (see the inset) is formed, as the pulse travels 5 mm.

Spectral evolution along the waveguide length is illustrated in Fig. 4. A few nonlinear interactive processes responsible for the formation of the supercontinuum can be seen. First, self-steepening of the optical pulse, associated with intensity-dependent group velocity [22], causes a sharp falling edge of the pulse. On the other hand, self-phase modulation produces blue-shifted spectral components at the falling edge, which walk-off very little relative to the edge, due to the low dispersion. Together with the self-steepening effect, these high-frequency components in turn help form a shaper edge, resulting in bluer shifts. Therefore, the flat and low dispersion triggers this positive feedback mechanism for optical shock formation and spectral broadening, which follows from A to B as shown in Fig. 4. Such a steep pulse edge transfers energy to a frequency range near 370 THz, ~230 THz away from the pulse carrier frequency. Second, tracking from B to C in Fig. 4, we note that, with accumulated dispersion, the falling edge becomes less steep, and newly generated blue-shifted frequencies are closer to the carrier, which improves the spectral flatness of the supercontinuum. Another effect of the dispersion is that the blue-shifted components walk away from the steep edge and overlap with the pulse tail, forming a beating pattern as shown in Fig. 3(b). Third, the pulse waveform in Fig. 3(b) has a high-power 'shoulder' at the beginning of its falling edge before the optical shock, which generates blue-shifted frequencies near the carrier following from A to D in Fig. 4. Fourth, third-harmonic generation occurs at 408 THz, and some frequency-resolved fringes are observed mainly due to cross-phase modulation by the input pulse [22]. The third-harmonic pulse sees a larger group delay and escapes from the envelope of the input pulse. This is why the third-harmonic spectrum becomes stable after a distance of ~1 mm.

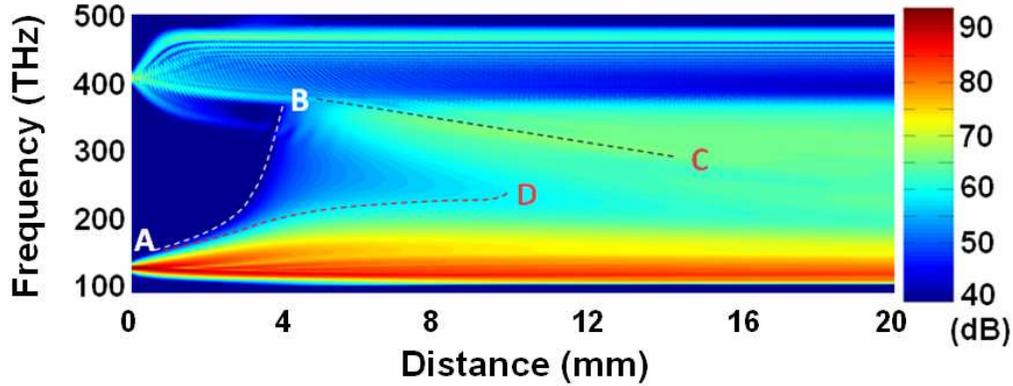

Fig. 4 Spectral evolution in the slot waveguides. Low dispersion causes dramatic spectral broadening and optical shock formation, from A to B. Then, accumulated dispersion makes the pulse falling edge less steep and improves spectral flatness, from B to C. Self-phase modulation produces blue-shifted frequencies near the carrier frequency, from A to D.

The dynamics of the self-steepening-induced supercontinuum generation, corresponding to Fig. 4, can be intuitively represented using spectrograms generated by the cross-correlation frequency-resolved optical gating (X-FROG) technique [2], in which an optical pulse is characterized simultaneously in time and frequency domains. As explained above, the fundamental pulse experiences dramatic self-steepening and spectral broadening in its propagation from 0 to 4 mm. Due to frequency-dependent group delay, the blue part of the edge walks off relative to the pulse, as seen at a distance of 10 mm, which forms a hockey-stick-like pattern in the spectrogram shown in Fig. 5.

The third-harmonic pulse exhibits more complex dynamics, and its evolution is significantly affected by group delay and pulse trapping induced by third harmonic generation [27, 28]. As shown in Fig. 5, the third-harmonic pulse is generated and cross-phase modulated at the beginning of propagation. From 1 mm to 2 mm, the tail of the third-harmonic pulse, which is after the steep edge of the fundamental pulse, walks away quickly, since its frequency is not located in the dispersion-flattened spectral range. The rest part of the pulse that coincides with the peak of the fundamental pulse is split into two parts. First, the low-frequency part travels slowly, and after it arrives at the steep edge of the fundamental pulse, it is blue-shifted due to cross-phase modulation and then escapes from the envelope of the fundamental pulse. Second, the high-frequency part is trapped by the fundamental pulse due to a nonlinear phase locking mechanism [27] and carries the dispersion property impressed by the fundamental pulse [27, 28], which is why its pattern in the spectrogram is also hockey-stick-like, although stretched 3 times in the frequency domain. As seen in Fig. 5, such phase-locked pulse trapping enables us to up-convert a 200-THz-wide supercontinuum that can be 2000 nm wide in wavelength, across a few-hundred-THz spectral region, to where a supercontinuum cannot be efficiently formed with an optical pulse at a local frequency, with flattened dispersion hardly achievable near material bandgap wavelength in practice.

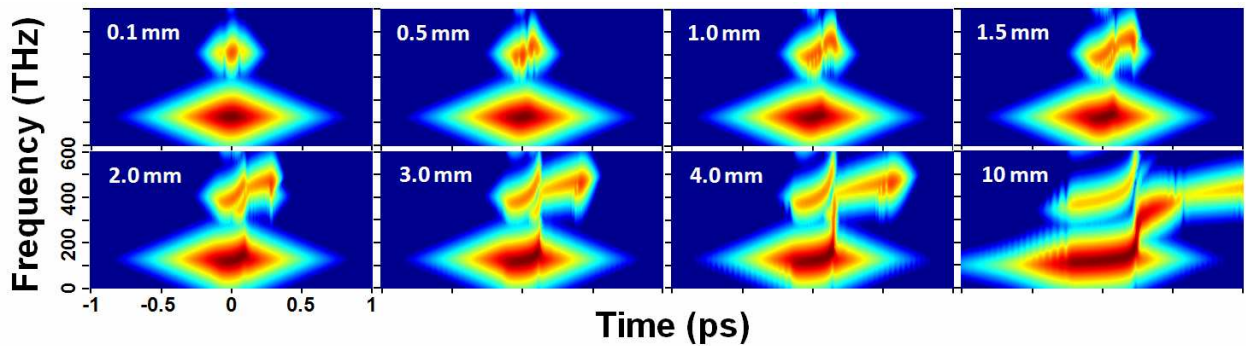

Fig. 5 Spectrogram evolution of the pulse. Strong self-steepening and spectral broadening occur in the pulse propagation from 0 to 4 mm. At 10 mm, the blue part of the pulse falling edge walks off due to dispersion. This forms a hockey-stick-like pattern in the spectrogram. A third-harmonic pulse is trapped by the fundamental pulse due to a nonlinear phase locking. Its spectrogram is also hockey-stick-like, though stretched 3 times in the frequency domain.

**Discussion**

Different from most of previously investigated supercontinua that were generated mainly due to either self-phase modulation in normal dispersion regime or high-order soliton fission and dispersion wave generation in anomalous dispersion regime [2], the supercontinuum reported here features a highly asymmetric spectrum caused by pulse self-steepening. Moreover, using a silicon nitride waveguide, one can have high output power and extended spectral range that are difficult to obtain in silicon waveguides.

**Conclusion**

In summary, we presented a dispersion tailoring technique that allows us to improve dispersion flatness by 30 times in integrated high-index-contrast waveguides. Benefiting from this, one can generate a two-octave supercontinuum on a chip by enhancing pulse self-steepening, which paves the way for ultrafast and ultra-wideband applications on an integrated nano-photonics platform. We believe that the on-chip supercontinuum generation would serve as a key enabler for building portable imaging, sensing, and positioning systems.